# Pre-stack and post-stack seismic inversion using quantum computing

Divakar Vashisth*[1,2], Rodney Lessard[2], and Tapan Mukerji[1]
[1]*Department of Energy Science and Engineering, Stanford University, USA*, [2]*SLB Software Technology Innovation Center, USA*

email: *divakar.vashisth98@gmail.com

## Abstract

Quantum computing harnesses the principles of quantum mechanics to solve problems that are intractable for classical computers. Quantum annealing, a specialized approach within quantum computing, is particularly effective for optimization tasks, as it leverages quantum tunneling to escape local minima and efficiently explore complex energy landscapes. In geosciences, many problems are framed as high-dimensional optimization problems, including seismic inversion, which aims to estimate subsurface impedances from seismic data for accurate geological interpretation and resource exploration. This study presents a novel application of quantum computing for seismic inversion, marking the first instance of inverting seismic data to estimate both P-wave and S-wave impedances using a quantum annealer. Building upon our prior work, which demonstrated the estimation of acoustic impedances from post-stack data using a two-step framework, we propose an enhanced workflow capable of inverting both post-stack and pre-stack seismic data in a single step. This advancement significantly reduces the number of qubits per model parameter (from 20 to 5) while improving computational speed (from 20 seconds to 6.3 seconds). The seismic inversion is implemented using the D-Wave Leap hybrid solver, achieving impedance estimation within 4–9 seconds, with the quantum processing unit (QPU) contributing just 0.043–0.085 seconds. Comparative analysis with simulated annealing reveals that quantum annealing produces impedance models closely matching true values in a single epoch, whereas simulated annealing requires 10 epochs for improved accuracy. These findings underscore the transformative potential of quantum computing for real-time, high-precision seismic inversion, marking a crucial step toward fully quantum-driven geophysical solutions.

# Introduction

Quantum computing marks a transformative shift in computational paradigms, offering promising avenues for addressing problems that are currently infeasible for classical computers. At the core of quantum computation lies the quantum bit, or qubit, which leverages fundamental quantum mechanical phenomena such as superposition and entanglement. These principles enable quantum systems to store and process information in fundamentally different ways compared to classical systems. Unlike classical bits, which are restricted to binary states of 0 or 1, qubits can exist in superpositions of these states, allowing quantum computers to handle multiple computations simultaneously. Additionally, entanglement creates a unique correlation between qubits, where the state of one qubit is intrinsically linked to the state of another, such that measuring the state of one qubit instantaneously determines the state of the other, regardless of the distance between them. This facilitates efficient and powerful information processing by enabling certain operations that are not possible in classical systems (Nielsen and Chuang, 2010). These quantum properties provide a significant computational advantage for specific tasks, exemplified by Shor's algorithm for integer factorization (Shor, 1994), which demonstrates an exponential speedup over the best-known classical algorithms, and Grover's algorithm for database searches (Grover, 1996), which achieves a quadratic speedup compared to classical brute-force search methods.

Quantum computers can be broadly categorized into two types: gate-based universal quantum computers, which utilize quantum logic gates to manipulate qubits for general-purpose quantum computing, and quantum annealers, which are specifically designed to solve optimization problems by finding the lowest energy state of a system. Quantum annealers leverage quantum phenomena such as tunneling and superposition to explore large solution spaces more efficiently than classical methods under certain conditions. For instance, the D-Wave quantum annealer has been applied to optimization problems such as traffic flow optimization (Neukart et al., 2017), protein folding simulations (Perdomo-Ortiz et al., 2012), and financial portfolio optimization (Rosenberg et al., 2016). Other studies, including Farhi et al. (2002) and Abel et al. (2021), have demonstrated the potential of quantum annealing to optimize target functions under specific scenarios, highlighting its advantages over classical alternatives for particular problem classes. Dukalski et al. (2023) estimated refraction residual statics for seismic data processing using quantum annealing.

Seismic inversion is a fundamental technique in geophysics, widely employed to derive quantitative rock property models from seismic survey data. It primarily focuses on estimating attributes like acoustic and elastic impedance, which are critical for characterizing subsurface formations (Tarantola, 2005; Sen and Stoffa, 2013). By interpreting seismic reflection amplitudes, this method refines our understanding of subsurface structures, enabling precise estimation of rock properties that are closely linked to hydrocarbon presence, type, and fluid saturation levels (Avseth et al., 2005; Grana et al., 2021). While seismic inversion is extensively utilized in hydrocarbon exploration, its applications extend beyond oil and gas. In the geothermal energy sector, it serves as a key tool for imaging subsurface structures with high resolution, aiding in the identification of geothermal reservoirs and fault zones (Krawczyk et al., 2019; Bredesen et al., 2020; Gao et al., 2021). Similarly, in Carbon Capture and Storage (CCS) initiatives, seismic inversion plays a crucial role in delineating and monitoring $CO_2$ storage sites. It ensures the long-term stability of sequestered carbon dioxide, mitigating potential environmental risks and supporting sustainable carbon management strategies (Arts et al., 2008; Liu et al., 2023; Hu et al., 2023).

In recent years, significant research efforts have focused on exploring the integration of quantum computing into seismic inversion. Cheng et al. (2022) implemented Amplitude Versus Offset (AVO) inversion to estimate elastic parameters using the hybrid quantum genetic algorithm (HQGA) and the quantum genetic algorithm (QGA). However, it is important to note that HQGA and QGA are quantum-inspired algorithms, not quantum computing methods. While they demonstrated superior performance compared to their classical counterparts, they operate entirely on classical hardware, merely adopting principles from quantum computing. This distinction is critical: quantum-inspired methods can enhance efficiency relative to traditional algorithms but do not fully harness the capabilities of quantum computing. In contrast, Greer and O'Malley (2020) attempted seismic inversion for a binary velocity model using the D-Wave quantum annealer. Albino et al. (2022) and Souza et al. (2022) reformulated the seismic inversion problem into a linearized traveltime inversion problem to estimate the slowness vector (inverse of velocity vector) in the equation $t_{ij} = 2\sum_{j=1}^{n} \frac{d_{ij}}{v_j} = 2\sum_{j=1}^{n} d_{ij} s_j$ , where $d_{ij}$, $v_j$ and $s_j$ are distance, velocity and slowness vectors respectively for $i^{th}$ source-receiver pair and $j^{th}$ layer.

The primary objective of this study is to address a practical, scalable, and business-relevant problem by estimating P- and S-wave impedances directly from acquired seismic trace data using a quantum computer. In our earlier work, Vashisth and Lessard (2024), we demonstrated the first application of seismic inversion on a quantum computer, employing the D-Wave quantum annealer to estimate acoustic impedances from post-stack seismic trace data through a two-step workflow (Figure 1). In the first step, the quantum annealer was used to estimate reflectivities from post-stack seismic data, followed by a second step where these estimated normal-incidence reflectivities were utilized to predict acoustic impedances using the same quantum technology. This study builds upon and significantly advances our previous framework (Vashisth and Lessard, 2024) by enabling the inversion of both post-stack and pre-stack seismic data in a single step, allowing for the direct estimation of P- and S-wave impedances from seismic data (Figure 2). This improvement eliminates the need to use the quantum annealer twice to solve the seismic inverse problem. The paper begins with a detailed overview of the seismic forward modeling framework used to compute seismograms from given P-wave velocity, S-wave velocity, and density models. This is followed by a description of the methodology for implementing seismic inversion on the quantum annealer to estimate P- and S-wave impedances. Subsequently, we apply the proposed methodology to both post-stack and pre-stack seismic inversion using examples from Das and Mukerji (2020), Grana et al. (2021), and Vashisth and Mukerji (2022). The results are then compared to those obtained using the simulated annealing algorithm. Furthermore, we discuss how the proposed framework represents a significant improvement over our earlier approach in Vashisth and Lessard (2024), where seismic inversion was implemented to estimate acoustic impedances using a quantum computer. Finally, we comment on the robustness of the proposed framework and reflect on the transformative potential of quantum processing units (QPUs) in advancing geophysical research and applications.

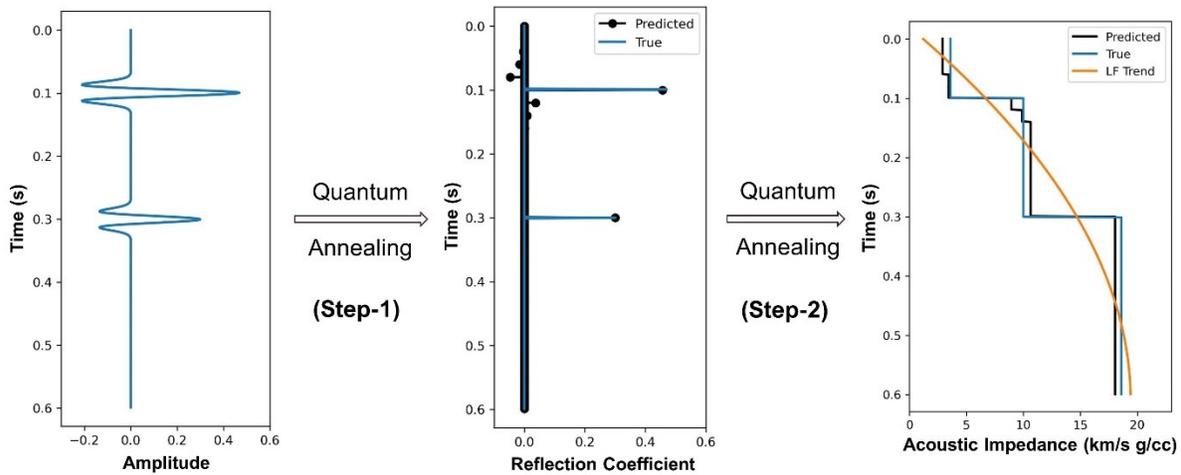

**Figure 1:** Two-step workflow proposed by Vashisth and Lessard (2024) for estimating acoustic impedances from post-stack seismic trace data. In the first step, reflection coefficients are estimated from seismic trace data using quantum annealing. In the second step, the normal-incidence reflectivities obtained in the first step are used to estimate acoustic impedances through quantum annealing.

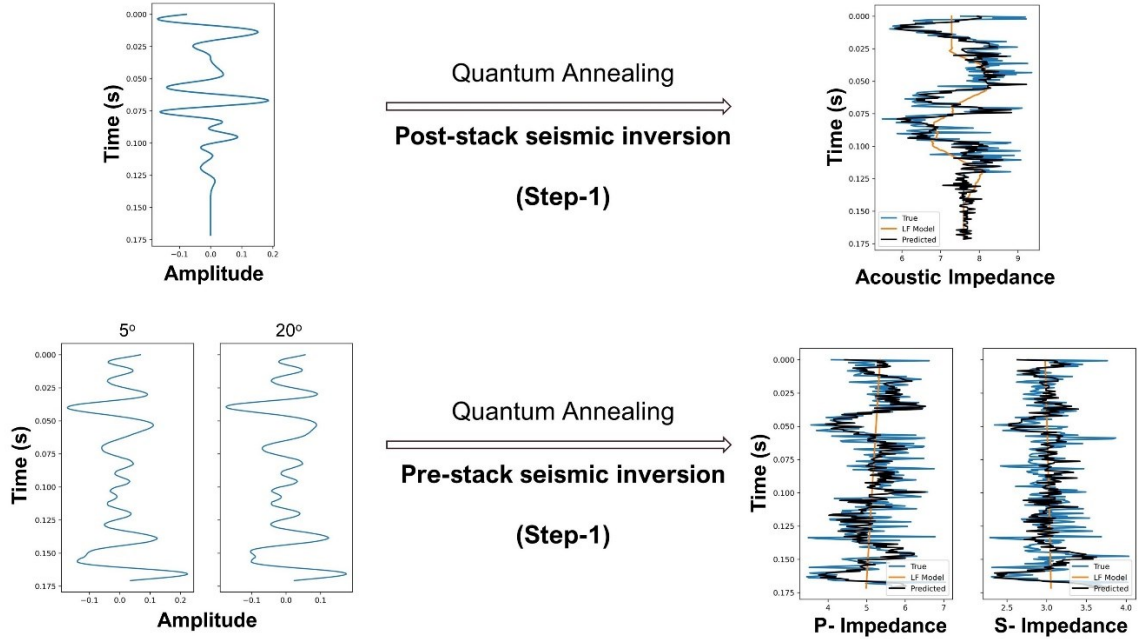

**Figure 2:** Proposed single-step workflow for directly estimating P- and S-wave impedances from both post-stack and pre-stack seismic trace data using quantum annealing.

## Seismic Forward Modeling

Seismic data can be understood as the result of a convolution process involving a source wavelet and a series of reflection coefficients, which represent the reflectivity associated with the elastic contrasts at geological layer boundaries. The reflection coefficients depend on the elastic properties of the layers, such as P-wave and S-wave velocities and density, which describe how seismic waves propagate through the subsurface. The seismic response at a specific time, considering the reflection angle, can be modeled using these properties.

Reflection Seismic Response

The seismic response at a particular two-way travel time $t$ and reflection angle $\theta$ is given by:

$$d(t,\theta) = w(t,\theta) * r_{PP}(t,\theta) = \int w(u,\theta)\, r_{PP}(t-u,\theta)du, \qquad (1)$$

where $w(t,\theta)$ is the source wavelet, and $r_{PP}(t,\theta)$ represents the PP-reflection coefficient series or reflectivity coefficients. These coefficients can be precisely calculated using the Zoeppritz equations, which describe the reflection and transmission of seismic waves at interfaces between different geological layers. For small reflection angles and weak elastic contrasts, these coefficients can be accurately approximated using linear relationships (Aki and Richards, 1980). We adopt this linear approximation based on amplitude variation with offset (AVO) attributes, as detailed in the works of Buland and Omre (2003) and Grana et al. (2021).

Amplitude Variation with Offset (AVO)

A common approach in seismic modeling is to use Amplitude Variation with Offset (AVO) attributes to approximate the reflection coefficients. The AVO attributes include the intercept $R$, the gradient $G$, and the curvature $F$, which are calculated using the differences in the elastic properties across the interface:

$$R = \frac{1}{2}\left(\frac{\Delta V_P}{\overline{V_P}} + \frac{\Delta \rho}{\overline{\rho}}\right),$$

$$G = \frac{1}{2}\frac{\Delta V_P}{\overline{V_P}} - 2\frac{\overline{V_S}^2}{\overline{V_P}^2}\left(\frac{\Delta \rho}{\overline{\rho}} + 2\frac{\Delta V_S}{\overline{V_S}}\right), \quad (2)$$

$$F = \frac{1}{2}\frac{\Delta V_P}{\overline{V_P}}$$

Here, $\Delta V_P$, $\Delta V_S$, and $\Delta \rho$ are the changes in P-wave velocity, S-wave velocity, and density across the interface, respectively, while $\overline{V_P}$, $\overline{V_S}$, and $\overline{\rho}$ are the average values of these properties at the interface.

Reflection Coefficient Expression

The reflection coefficient $r_{PP}(\theta)$ at an angle $\theta$ can be expressed as:

$$r_{PP}(\theta) = R + G\sin^2\theta + F(\tan^2\theta - \sin^2\theta), \quad (3)$$

where the reflection angle $\theta$ is the angle between the incident seismic wave and the normal to the reflecting interface.

This expression can be further expanded using the elastic properties to derive a more detailed formula, highlighting the influence of changes in P-wave velocity, S-wave velocity, and density on the reflection coefficient.

Time-Continuous Reflectivity

The reflection coefficient can be extended to a continuous reflectivity function over time (Stolt and Weglein, 1985):

$$c_{PP}(t,\theta) = c_P(\theta)\frac{\partial}{\partial t}\ln V_P(t) + c_S(\theta)\frac{\partial}{\partial t}\ln V_S(t) + c_\rho(\theta)\frac{\partial}{\partial t}\ln \rho(t), \quad (4)$$

$$c_P(\theta) = \frac{1}{2}(1 + tan^2\theta),$$

$$c_S(\theta) = -4\frac{\overline{V_S}^2}{\overline{V_P}^2}sin^2\theta, \quad (5)$$

$$c_\rho(\theta) = \frac{1}{2}\left(1 - 4\frac{\overline{V_S}^2}{\overline{V_P}^2}sin^2\theta\right)$$

where $c_P(\theta)$, $c_S(\theta)$, and $c_\rho(\theta)$ are coefficients dependent on the reflection angle $\theta$ and the average velocities of the seismic waves. These coefficients dictate how changes in the velocities and density with respect to time affect the reflectivity.

Given knowledge of the elastic properties of the subsurface layers, the seismic response can be calculated as a convolution of the source wavelet with this continuous reflectivity function:

$$d(t, \theta) = w(t, \theta) * c_{PP}(t, \theta), \quad (6)$$

Discrete Seismic Forward Model

In a practical scenario, the model variables are often discretized into a vector $m = [\ln V_P, \ln V_S, \ln \rho]^T$, where each component vector includes time samples representing the elastic properties at different layers. The continuous reflectivity series is also discretized, leading to the following relationship:

$$c = ADm = Am', \quad (7)$$

where $A$ is a block matrix that encodes the time samples of the reflectivity coefficients, $D$ is a first-order differential matrix, and $m'$ is the time derivative of the model parameters.

The seismic response $d$ can then be modeled as a discrete convolution:

$$d = WAm' + e, \quad (8)$$

where $W$ is a block-diagonal matrix containing the discretized wavelets, and $e$ represents the data error.

Seismic Data Representation

The matrix-vector multiplication used in this model can be made more explicit by separating the time and angle dependence:

$$d(t_i, \theta_i) = W(\theta_i)A(t_i, \theta_i)m'(t_i) + e(t_i, \theta_i), \quad (9)$$

where $W(\theta_i)$ represents the wavelet associated with the reflection angle $\theta_i$, and the vector $d$ contains the seismic data corresponding to different angles and times.

## Framing Seismic Inversion as an Optimization Problem

To solve the post-stack and pre-stack seismic inverse problem on the quantum annealer, we frame it as an optimization problem. The objective function that we aim to optimize or minimize through Quantum Annealing is represented by:

$$E = \sum_{t,\theta} \left(d_{seis}(t,\theta) - W(\theta)A(t,\theta)m'_{pred}(t)\right)^2 + \sum_{t} \lambda \left(m_{pred}(t) - m_{LF}(t)\right)^2 \quad (10)$$

where $d_{seis}(t,\theta)$ is the recorded or observed seismic data, $W(\theta)$ is the source wavelet associated with the reflection angle $\theta$, $m'_{pred}(t) = Dm_{pred}(t)$ where $D$ is a first-order differential matrix, $m_{pred}(t) = \left[\ln V_{P\,pred}(t), \ln V_{S\,pred}(t), \ln \rho_{pred}(t)\right]^T$, and $m_{LF}(t) = [\ln V_{P\,LF}(t), \ln V_{S\,LF}(t), \ln \rho_{LF}(t)]^T$ is the background or low-frequency trend of the model parameters known a priori. $\lambda$ is the regularization parameter. Hence, the objective is to optimize equation (10) using a quantum computer to determine $m_{pred}(t)$. Once the optimal solution is obtained, substituting $m_{pred}(t)$ into equation (8) should yield seismograms that closely align with the observed (or "true") seismic data. In seismic inversion, a low-frequency model is often incorporated (in this case, as a regularization term) to compensate for the lack of low-frequency information in the seismic data. This low-frequency model helps stabilize the inversion process and ensures that the solution remains physically meaningful. Without it, seismic inversion might produce non-unique or unrealistic results. Therefore, integrating a low-frequency model derived from well logs or other independent sources into the seismic inversion process is critical for accurate subsurface characterization.

To optimize equation (10), we leverage the D-Wave quantum annealer, which is adept at solving problems expressible as Quadratic Unconstrained Binary Optimization (QUBO) or Ising model Hamiltonian. In the quantum annealing paradigm, computations are encoded as finding the ground state of an Ising model Hamiltonian:

$$H(\sigma) = \sum_{ij} \sigma_i J_{ij} \sigma_j + \sum_{i} \sigma_i h_i \quad (11)$$

where $\sigma_i$ represents the spin variables taking values $\pm 1$, $h_i$ represents the energies (biases) of individual spins, and $J_{ij}$ represents the coupling/interaction strengths. Hence, to optimize equation (10) using a quantum annealer, it must first be reformulated as a minimization problem expressed in terms of the Ising Hamiltonian $H$ in equation (11). Criado and Spannowsky (2023) introduced a framework for encoding differential equations as the ground state of an Ising model Hamiltonian. Equation (10) can be rewritten as:

$$E(m) = d^T d - 2d^T WADm + m^T(D^T A^T W^T WAD + \lambda I)m - 2\lambda m_{LF}^T m + \lambda m_{LF}^T m_{LF} \quad (12)$$

where, $d$ corresponds to $d_{seis}(t,\theta)$ matrix, $m$ corresponds to $m_{pred}(t)$ matrix of equation (10), and $I$ represents the identity matrix. We can also write equation (12) as:

$$E(m) = m^T(D^TA^TW^TWAD + \lambda I)m - (2d^TWAD + 2\lambda m_{LF}^T)m + const. \qquad (13)$$

So, the loss function (equation 13) now includes quadratic terms involving $m^TQm$, where $Q = (D^TA^TW^TWAD + \lambda I)$, and linear terms involving $bm$, where $b = -(2d^TWAD + 2\lambda m_{LF}^T)$. The constant terms $(d^Td + \lambda m_{LF}^T m_{LF})$ can be ignored since they do not affect optimization. The final step to optimize equation (13) on a quantum annealer is the binary encoding of each weight ($m$ in our case) using spin variables. Since elastic parameters are real numbers rather than binary values, multiple qubits are required to represent each weight accurately. Each component of the weight matrix ($w_i$) can be represented using binary spin variables (Criado and Spannowsky, 2023):

$$w_i = c_i + s_i \sum_{\alpha=1}^{n_{spins}} \sigma_i^{(\alpha)} 2^{-\alpha} \qquad (14)$$

where $c_i$ is the center value for each parameter, $s_i$ is the scaling factor for each parameter, $\sigma_i^{(\alpha)}$ are binary spin variables ($\pm 1$). For an iterative algorithm, $c_i$ is updated iteratively with the best results from the previous run ($c_i^{(k+1)} = w_i^{(k)}$), and $s_i$ is scaled down iteratively by a factor of $S$ for finer adjustments ($s_i^{(k+1)} = Ss_i^{(k)}$). In our implementation, five qubits are used to represent each weight ($m$), $c_i$ is substituted with $m_{LF}(t)$, and $s_i$ is set to 0.1. Consequently, equation (14) replaces $m$ with binary spin variables ($\sigma_i$) in the loss function (equation 13). Each model parameter $[\ln V_P, \ln V_S, \ln \rho]$ at every time step is encoded using five spin variables. After reformulating equation (10) as an Ising model Hamiltonian (equation 11), the optimization is performed using the D-Wave quantum annealer (Figure 3). While optimization problems are typically solved iteratively, where the solution of one iteration guides the next until convergence is reached, the D-Wave quantum annealer's ability to handle Ising model Hamiltonians allowed for convergence to the optimal solution in just one iteration, eliminating the need for multiple iterations.

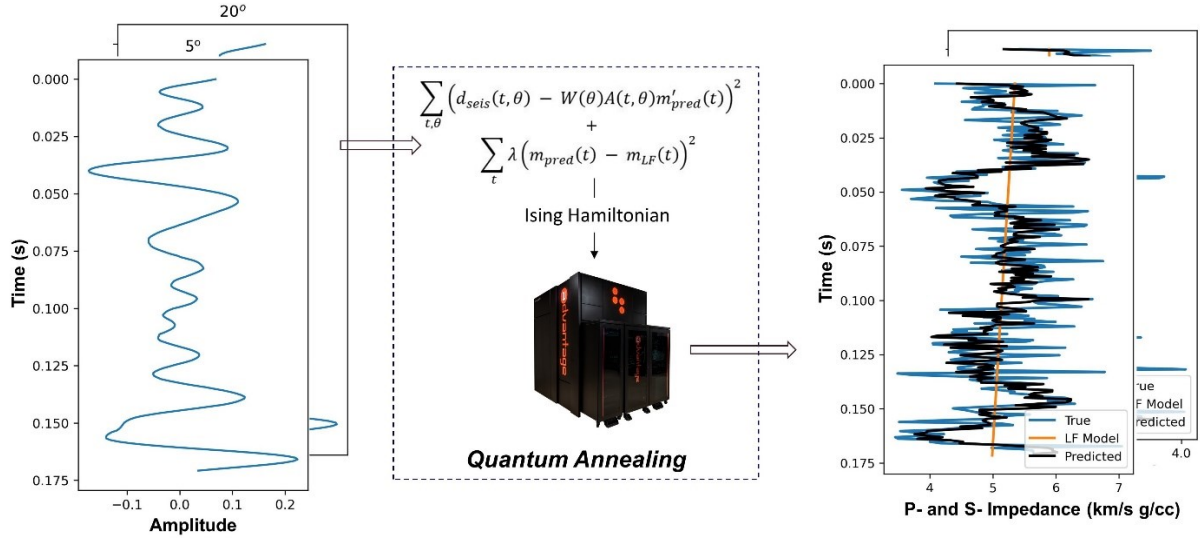

**Figure 3:** Single-step workflow for estimating P- and S-wave impedances from seismic data using quantum annealing, where the seismic inversion objective function is encoded as the ground state of an Ising Hamiltonian. The quantum annealer minimizes the objective function to output the P- and S-wave impedances.

## Results

We begin with the inversion of pre-stack seismic data for three partial angle stacks: near (12º), mid (24º), and far (36º), using the dataset from Grana et al. (2021), as illustrated in Figure 4. The estimated P-impedance and S-impedance profiles from the quantum annealer are shown in Figure 5. The predicted P- and S-wave impedances closely align with the true impedance models, yielding root-mean-square (RMS) errors of 1.320 and 0.769 km/s·g/cc, respectively. To further emphasize the effectiveness of the quantum computing approach in addressing the seismic inverse problem, we compared these results with those obtained using the well-known simulated annealing algorithm. The simulated annealing algorithm followed an identical workflow and aimed to estimate the same model parameters ($m_{pred}(t)$ of equation 10), mirroring the objective of quantum annealing. The results from the simulated annealing algorithm are also presented in Figure 5. Notably, the simulated annealing algorithm did not produce satisfactory results in a single iteration, with RMS errors of 1.451 and 0.952 km/s·g/cc for P- and S-wave impedances, respectively. To improve accuracy, we employed a multi-iterative approach, running the simulated annealing algorithm for 10 epochs. This extended approach improved the results, achieving RMS errors of 1.168 and 0.740 km/s·g/cc. However, running the quantum annealer for an additional epoch produced results similar in accuracy to the 10-epoch simulated annealing output. The seismic inverse problem on the D-Wave quantum annealer could be approached in two ways: (1) direct computation on QPUs or (2) using a hybrid quantum-classical solver. While direct QPU computation would ideally be the most efficient approach, the current hardware limitations, specifically the insufficient number of fully connected qubits, necessitated the use of the hybrid solver. The D-Wave Leap Hybrid solver combines the strengths of classical computing and quantum annealing to solve complex optimization problems. The hybrid solver decomposes large problems into smaller sub-problems, each fitting within the constraints of the quantum annealer. These sub-problems are processed on the QPU, leveraging quantum mechanics to search for optimal solutions. The partial quantum solutions are then returned to the classical system, where classical algorithms refine and integrate them into a final, optimized result. This hybrid approach allows for greater flexibility in tackling a wide range of problems, scalability for problems too large for direct quantum computation, and enhanced noise tolerance through classical correction techniques. Therefore, Figure 5 presents the seismic inversion results derived from the hybrid solver, showcasing the estimation of P- and S-wave impedances from pre-stack seismic data.

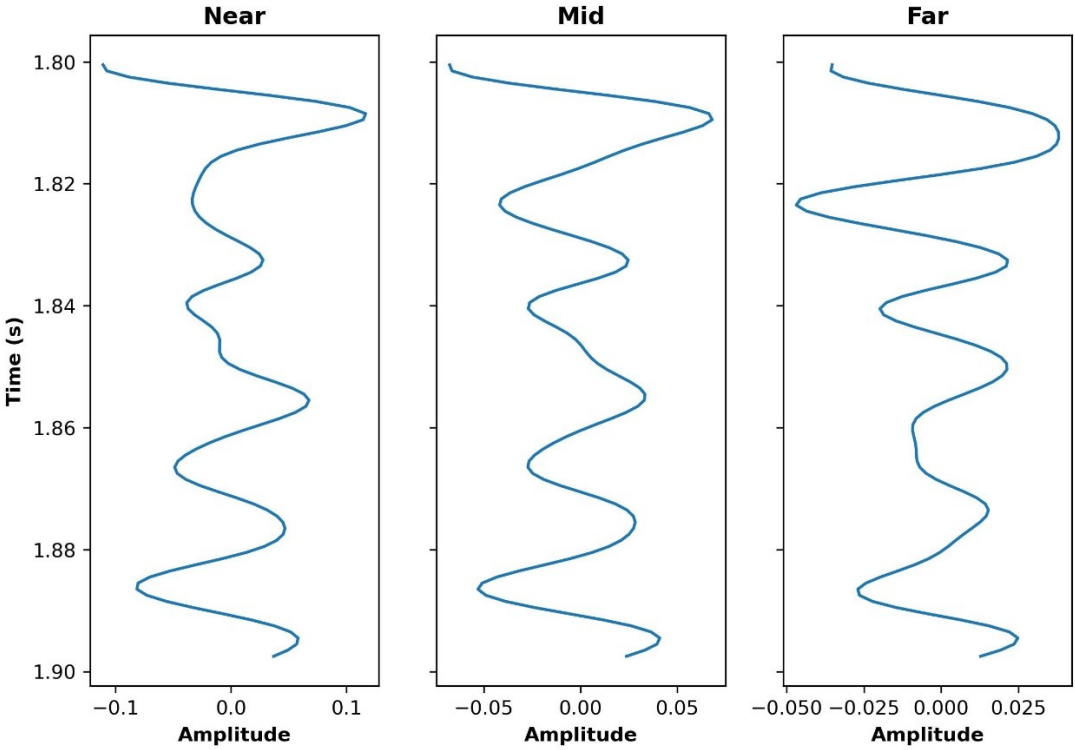

**Figure 4:** Pre-stack seismic data consisting of three partial angle stacks: near (12º), mid (24º), and far (36º), as presented in Grana et al. (2021).

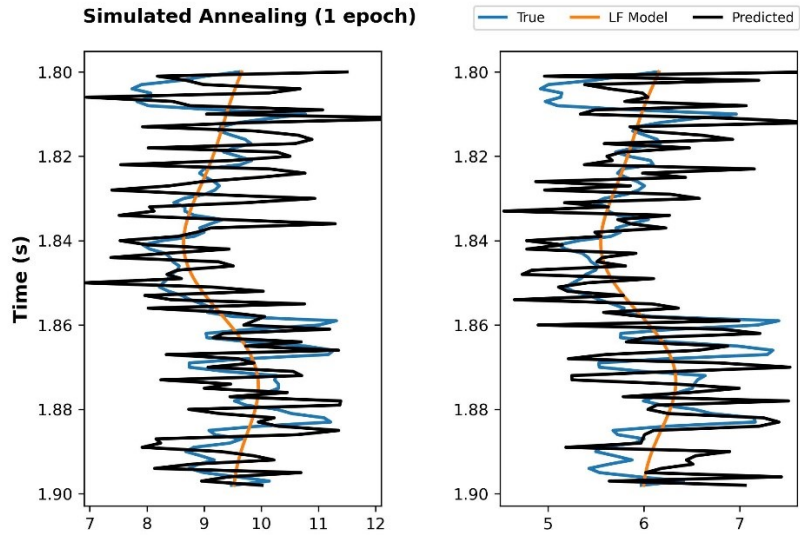
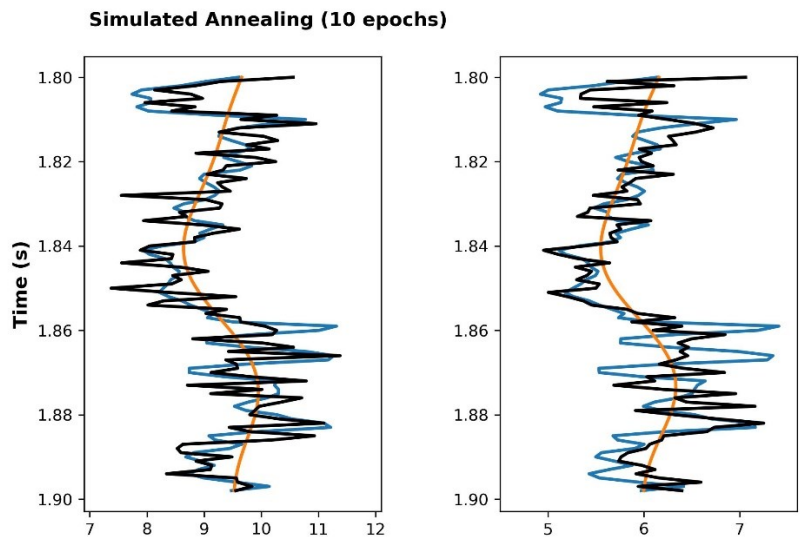
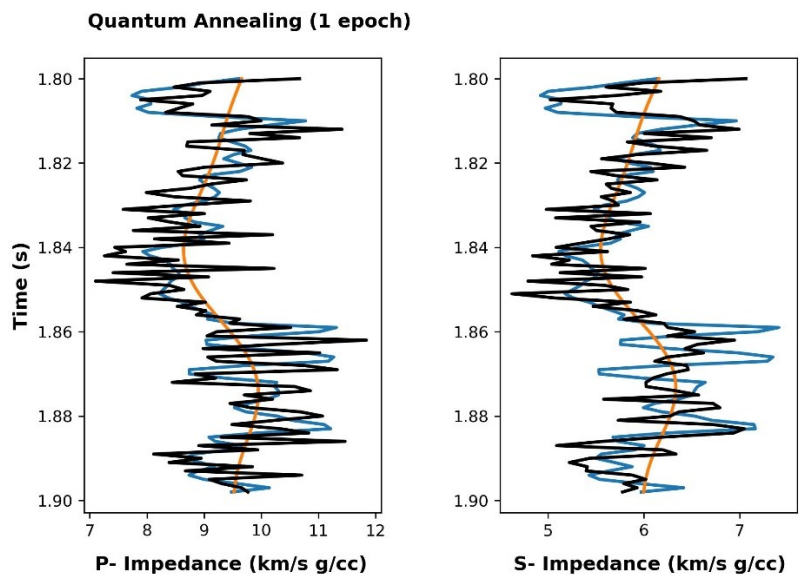

**Figure 5:** Comparison of P- and S-wave impedance profiles estimated using simulated annealing and quantum annealing for the pre-stack seismic data shown in Figure 4.

Figure 6 presents the pre-stack seismic inversion results for angle gathers at incident angles of 5° and 20°, obtained using the quantum annealer for selected examples from Das and Mukerji (2020). Unlike the original study, which focused on predicting petrophysical properties such as porosity and clay volume, our research aims to estimate elastic properties, specifically P-wave and S-wave impedances. The P- and S-wave impedances predicted by the quantum annealer closely align with the true impedance models. For the second example (Figure 6), the RMS errors for P- and S-wave impedances are 0.621 and 0.286 km/s·g/cc, respectively. Similarly, for the third example (Figure 6), the RMS errors are 0.582 and 0.297 km/s·g/cc, respectively. Our analysis was not confined to pre-stack seismic inversion alone. To further validate the efficacy of the proposed workflow, we also conduct post-stack seismic inversion on a set of examples from Vashisth and Mukerji (2022). These examples were also previously employed in Vashisth and Lessard (2024) to showcase the application of quantum annealing for post-stack seismic inversion. The results of the post-stack inversion using the quantum annealer are shown in Figure 7. The strong agreement between the predicted and true acoustic impedance models underscores the reliability and accuracy of our proposed framework. Table 1 summarizes the RMS errors between the true and estimated acoustic impedance profiles obtained using both the simulated annealing algorithm and the quantum annealer.

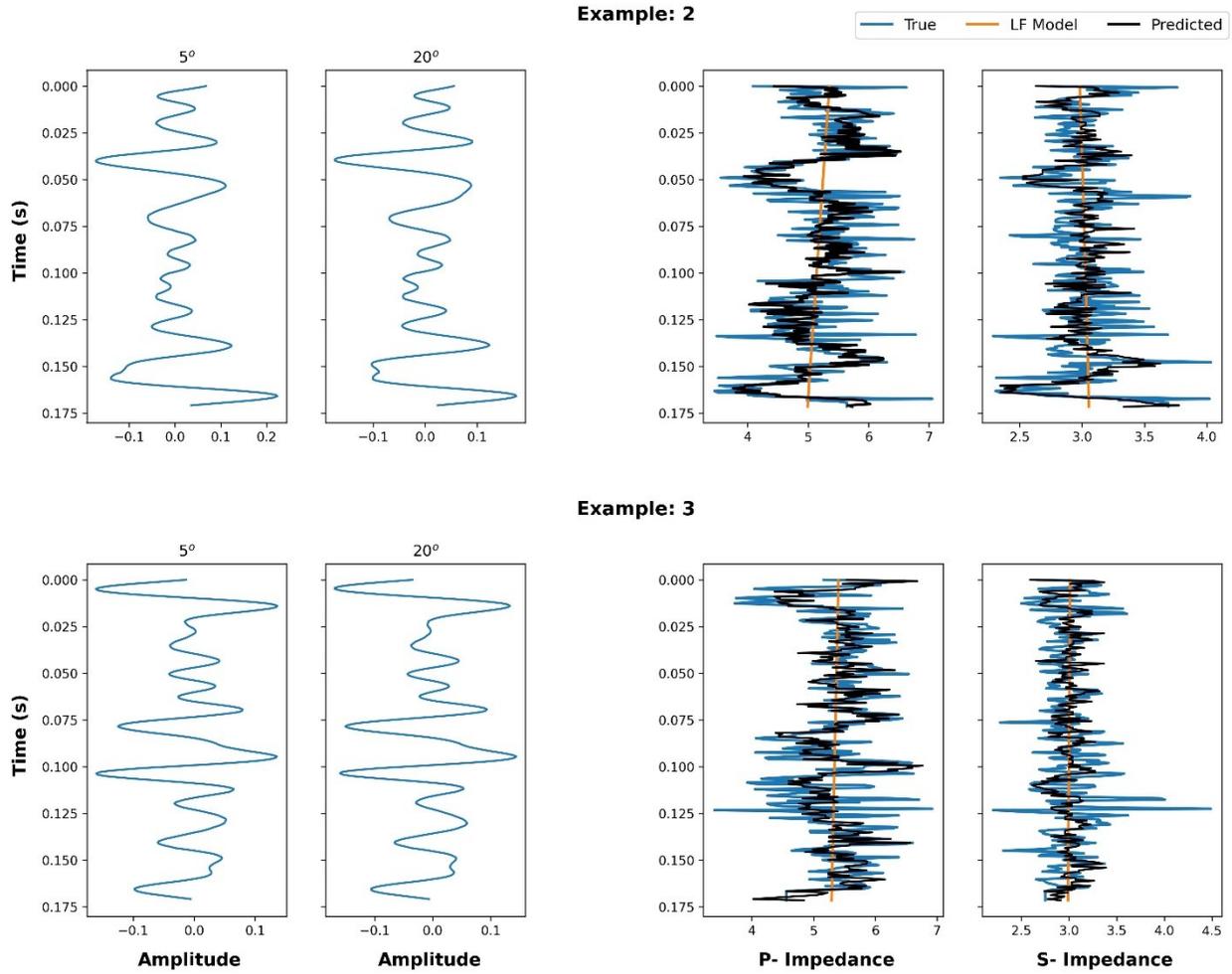

**Figure 6:** Pre-stack seismic inversion results using the quantum annealer for examples from Das and Mukerji (2020). The input seismic traces are shown on the left, while the output P- and S-wave impedance profiles are displayed on the right.

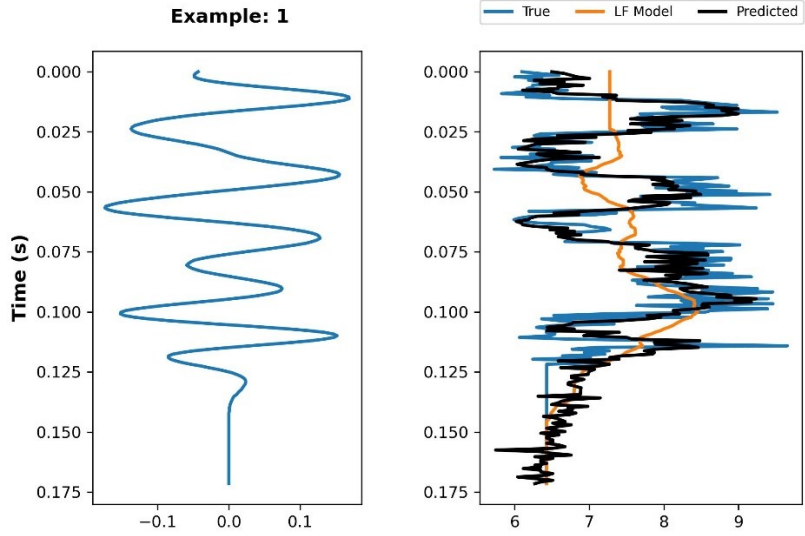
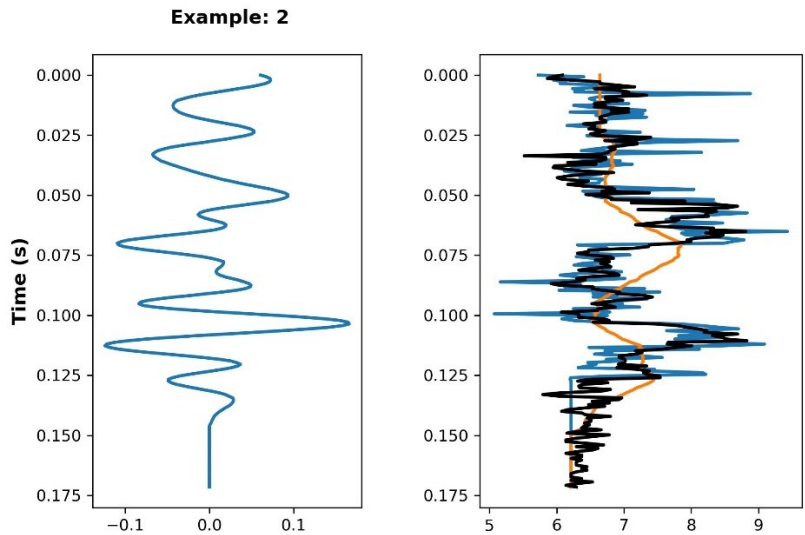
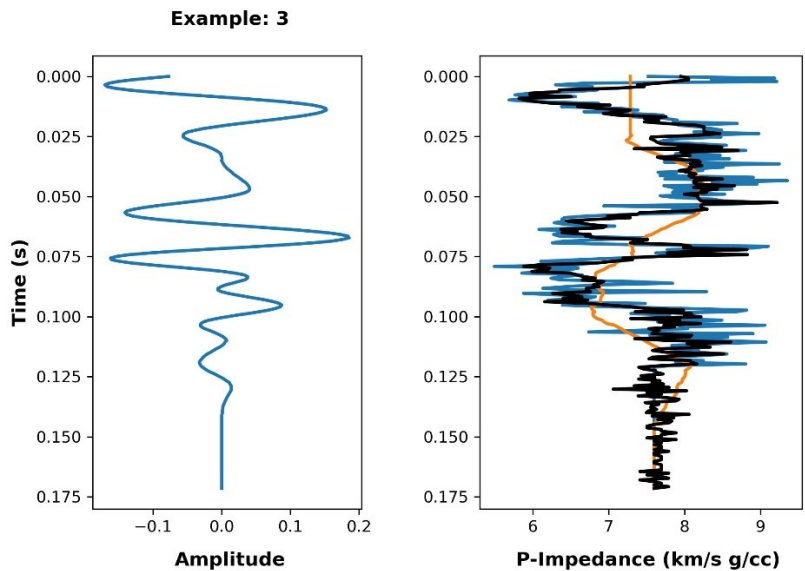

**Figure 7:** Post-stack seismic inversion results using the quantum annealer for examples from Vashisth and Mukerji (2022). The input seismic traces are shown on the left, with the output acoustic impedance profiles presented on the right.

**Table 1:** RMS errors (km/s·g/cc) between the true and estimated acoustic impedance profiles obtained using quantum annealing and simulated annealing algorithms for the three post-stack seismic examples shown in Figure 7.

| Annealer | Example 1 | Example 2 | Example 3 |
|---|---|---|---|
| **Simulated** | 0.671 | 0.625 | 0.711 |
| **Quantum** | 0.559 | 0.507 | 0.471 |

Vashisth and Lessard (2024) introduced a two-step workflow for estimating acoustic impedances from post-stack seismic data using a quantum annealer. In this study, we significantly enhance that approach by enabling the estimation of subsurface impedances from both post-stack and pre-stack seismic data in a single step. This improvement allows for the inversion of a larger number of model parameters ($m_{pred}(t)$ from Equation 10) using fewer spins (qubits) per parameter (5 instead of 20) and achieves results at a substantially faster computational speed (6.3 seconds compared to 20 seconds). Figure 8 presents a comparative analysis of the results from the previous workflow ("Old") proposed by Vashisth and Lessard (2024) and the enhanced framework ("New") developed in this study, using the same post-stack seismic example (Example 1 from Figure 7). The results clearly demonstrate the improved accuracy of our proposed methodology, with an RMS error of 0.559 km/s·g/cc, compared to 0.781 km/s·g/cc achieved by the earlier workflow. The reported computational time of 6.3 seconds for estimating acoustic impedances using the quantum annealer refers to the total runtime of the hybrid solver, which includes both the classical processing time and the time spent on the QPU. Remarkably, within the 6.3 seconds, the QPU execution time accounts for only 0.085 seconds, with the remainder attributed to the classical component of the hybrid solver. Similarly, for the pre-stack examples, the total runtime of the hybrid solver is 4 seconds for the first example (Grana et al., 2021) and 9 seconds for the second and third examples (Das and Mukerji, 2020), with corresponding QPU times of 0.043 and 0.085 seconds, respectively. It is worth noting that increasing the number of spins per variable generally improves the accuracy of the estimated impedances. However, this improvement comes at the cost of increased total runtime on the annealer. The billing model for quantum annealers charges based on total runtime, not just QPU execution time, making it essential to carefully balance computational accuracy and cost-efficiency. Figure 9 illustrates the variation of total runtime, QPU time, objective function loss (from Equation 10), and RMS loss of predicted impedances with respect to the number of spins (qubits) per weight ($m_{pred}(t)$ from Equation 10) for the post-stack and pre-stack case studies. Across all case studies presented in this paper, we use 5 spins per model parameter to strike a balance between computational efficiency and objective function accuracy.

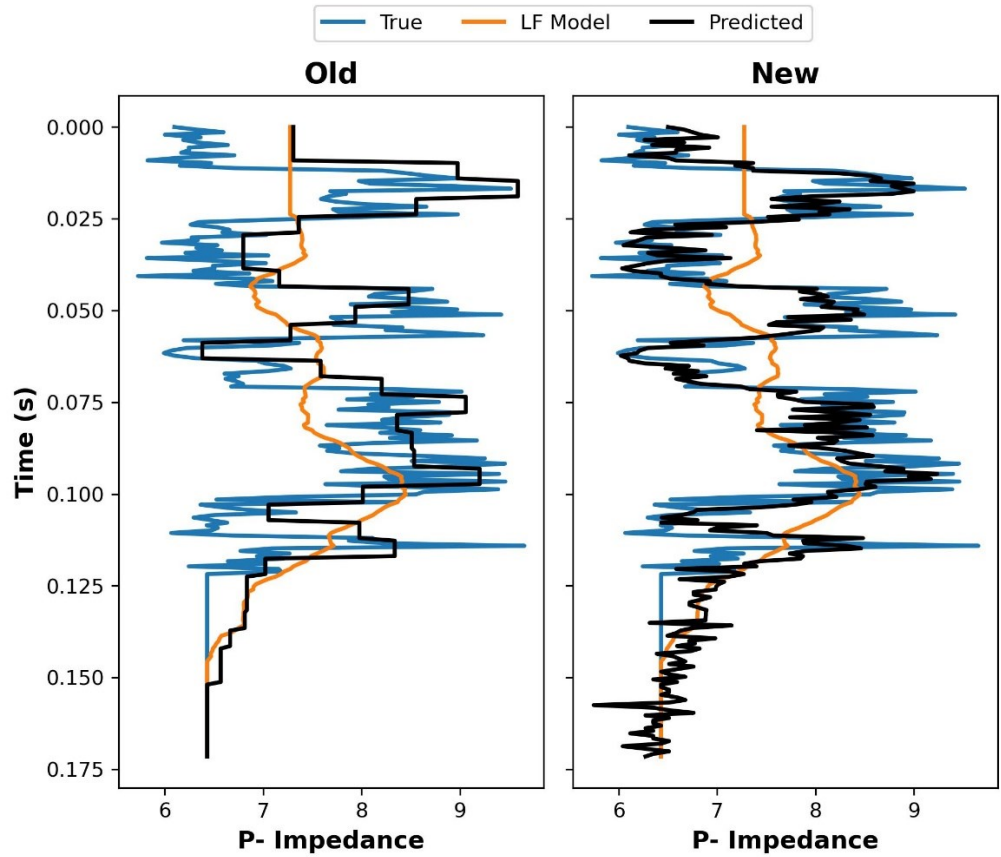

**Figure 8:** Comparison of acoustic impedance profiles estimated using the two-step workflow proposed by Vashisth and Lessard (2024) ("Old") and the improved single-step workflow introduced in this study ("New") for the same post-stack seismic example.

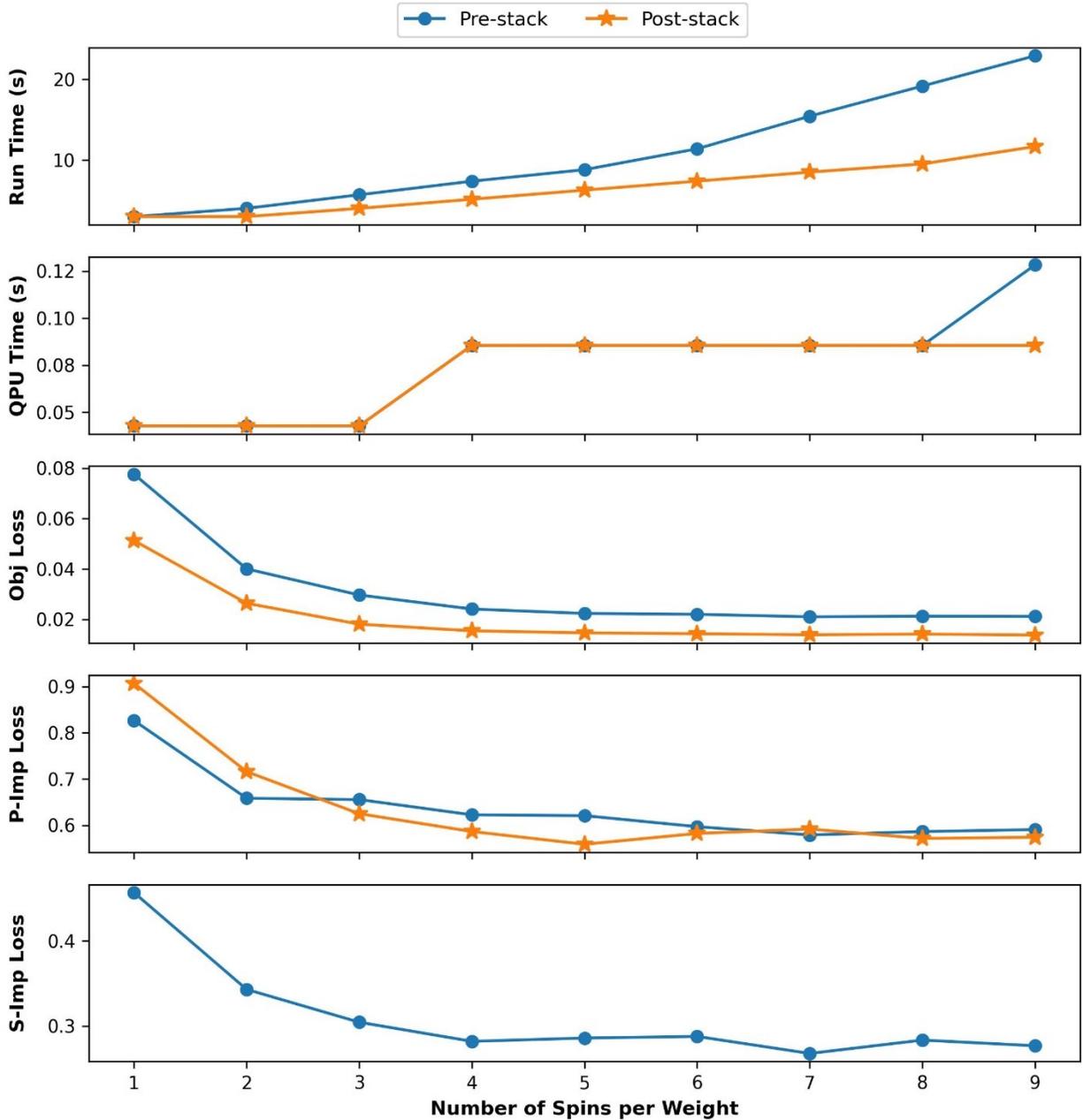

**Figure 9:** Variation of total runtime, QPU time, objective function loss (Equation 10), and RMS loss of the predicted P- and S- impedance profiles with respect to the number of spins (qubits) per weight ($m_{pred}(t)$) for the post-stack and pre-stack seismic inversion case studies.

## Conclusions

In this study, we have implemented both post-stack and pre-stack seismic inversion using quantum computing to estimate P- and S-wave impedances from seismic trace data. For both scenarios, we have inverted three examples each, and the impedances predicted by the quantum computer closely align with the true impedance models. The D-Wave Leap hybrid solver required only 4–9 seconds to estimate impedances, with the quantum processing unit (QPU) accounting for just 0.043–0.085 seconds of this runtime, while the remainder was spent on the classical component of the hybrid solver. These results demonstrate the potential of QPUs to address seismic inversion problems in under a second. To the best of our knowledge, this is the first time where seismic inversion has been implemented on a quantum computer to estimate both P- and S-wave impedances. Previously, Vashisth and Lessard (2024) pioneered the use of a quantum computer to estimate acoustic impedances from post-stack seismic data. Building upon that foundation, we have significantly improved the workflow by enabling the inversion of both post-stack and pre-stack seismic data in a single step-eliminating the need for two steps. This enhancement allows us to predict a larger number of model parameters using fewer spins (5 spins per parameter compared to 20) and achieves results at a substantially faster speed (6.3 seconds compared to 20 seconds). To further emphasize the effectiveness of our quantum computing approach, we compared the impedances estimated using the quantum annealer with those predicted by the simulated annealing algorithm. Both approaches followed an identical workflow and aimed to estimate the same set of model parameters. Remarkably, the quantum annealer predicted impedances closely matched the true models in just a single epoch, whereas simulated annealing required 10 epochs to achieve better accuracy. Additionally, running the hybrid solver for one more iteration resulted in predictions similar in accuracy to those obtained from the 10-epoch simulated annealing algorithm. We opted for the hybrid quantum-classical approach over direct QPU processing due to the current limitations in qubit connectivity on existing quantum processors. However, our proposed workflow for estimating impedances from seismic data is not only robust but also highly adaptable, with the exact procedure for the hybrid solver capable of being fully executed on QPUs as they continue to advance. With ongoing developments in quantum annealing technology, such as the increasing qubit count and connectivity, we anticipate that in the near future, we will have access to the requisite number of fully connected qubits to efficiently solve the seismic inverse problem. Furthermore, the proposed workflow is versatile and can be readily adapted to frame other optimization problems relevant to geosciences as Ising Hamiltonians. This adaptability ensures efficient implementation on quantum annealers, paving the way for broader applications of quantum computing in solving complex geoscience problems.

## Acknowledgements

We acknowledge the SLB Software Technology Innovation Center (STIC) and the sponsors of the Stanford Center for Earth Resources Forecasting (SCERF) for their support in conducting this research.